\documentclass[onecolumn,letterpaper,amsmath,amssymb,floatfix,aps,superscriptaddress]{revtex4}
\usepackage{hyperref}
\usepackage{graphicx}
\usepackage{dcolumn}
\usepackage{bm}
\usepackage{epsfig,color}
\usepackage{wasysym}

\def\ul#1#2{\textstyle{\frac#1#2}}

\begin{document}

\title{Polymers pushing Polymers:\\
Polymer Mixtures in Thermodynamic Equilibrium with a Pore}
\author{R. Podgornik$^{(a,b)}$, J. Hopkins$^{(a)}$, V.A. Parsegian$^{(a)}$ and M. Muthukumar$^{(a,c)}$\\
~\\
\sl 
${(a)}$ Department of Physics, University of Massachusetts, Amherst MA 01003, USA\\
${(b)}$ Department of Physics, Faculty of Mathematics and Physics, University of Ljubljana, and Department of Theoretical Physics, J. Stefan Institute, 1000 Ljubljana, Slovenia\\
${(c)}$ Polymer Science and Engineering Department, University of Massachusetts, Amherst, MA 01003, USA}

\begin{abstract} 
{\bf Abstract}
We investigate polymer partitioning from polymer mixtures into nanometer size cavities by formulating an equation of state for a binary polymer mixture assuming that only one (smaller) of the two polymer components can penetrate the cavity. Deriving the partitioning equilibrium equations and solving them numerically allows us to introduce the concept of "{\sl polymers-pushing-polymers}" for the action of non-penetrating polymers on the partitioning of the penetrating polymers.  Polymer partitioning into a pore even within a very simple model of a binary polymer mixture is shown to depend in a complicated way on the composition of the polymer mixture and/or the pore-penetration penalty. This can lead to enhanced as well as diminished partitioning, due to two separate energy scales that we analyse in detail.
\end{abstract}

\maketitle

\section{Introduction}

First analyzed within the framework of size exclusion chromatography \cite{Teraoka} passive partitioning of polymers into nanoscale cavities has gained much broader relevance \cite{Muthubook}. Active forcing of polymers into nanosize pores underlies some applications of the "osmotic stress" technique,  devised originally to probe inter- and intramolecular forces \cite{OST} but also used successfully to probe changes in the size of nano-cavities of some proteins \cite{OST1}. Size-dependent partitioning of water-soluble polymers is particularly important for the sizing and probing of water-permeable ion channels \cite{Krasilnikov,Krasilnikov2,Krasilnikov3} and for single-molecule mass spectroscopy \cite{J2K,J2K2,J2K3}. In addition, forcing polymers into nanosize tubes is relevant to recent studies of controlled ejection of viral genomes from capsids \cite{Gelbart}, where osmotic stress is applied to push a DNA molecule into or to allow it out of the viral capsid by the action of PEG (poly-ethylene-glycol) dissolved in the ambient solution \cite{Evilevitch}. In this last case, the identities of the polymer forced into the nano-cavity (DNA) and the polymer pushing it (PEG) differ. 

\begin{figure}[h!]
\centering
\includegraphics[width=10cm]{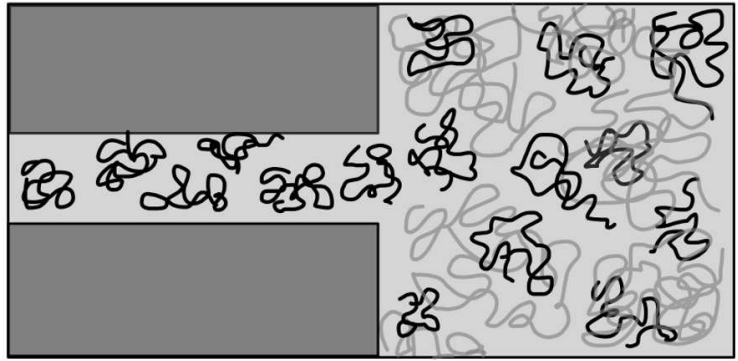}
\caption{A polymer solution composed of two polymer types (black and grey) in equilibrium with a pore. Only one component (black) of this binary polymer mixture is allowed to enter the pore, with a energy penalty $\Delta f$. We assume throughout that the two polymer species differ only in length. \label{schematic}}
\end{figure}

We analyze a variation on the problem. Here the polymer solution is heterogeneous, composed of various sizes of the (same type of) polymers, for which the accessibility of the pore varies depending on their size. We thus assume that the polymer solution is polydisperse and that of the different size polymer chains only a single type can enter the pore (Fig. \ref{schematic}). We show that in this case the polymers that are too big, or are otherwise prevented from entering the pore, act to push the other polymer into the pore.

Usually penetration of polymers into molecular nano-pores incurs an energy penalty. It depends also on the concentration of polymer chains outside the pore, assumed to be in equilibrium with the pore. Several approaches have been pursued in the theoretical elucidation of this problem \cite{Daoud,Sergey,Teraoka}. The main difference between previous approaches and the one advocated here is that we start with a consistent free energy of the polymer-solvent mixture that allows us to formulate the thermodynamic equilibrium for the pore-external solution system in terms of the chemical potential of the polymer species as well as the chemical potential (osmotic pressure) of the solvent. This is important. We want to generalize the formalism in such a way that we will be able to deal with a mixture of polymers, whose components are selectively allowed to enter the pore. We introduce a simple {\sl ansatz} for the free energy of a mixture of (the same type of) polymers varying only in their sizes, consistent with our previous phenomenological fit to the equation of state of bulk uncharged polymers \cite{EOS}. Calculating the partition coefficient allows us to ascertain that indeed some polymers can push others to enter the pore in a kind of osmotic tug-of-war.

We build our analysis around the most simple case of a PEG polymer solution and first review the properties of a monodisperse solution. We then derive its equation of state, consistent with empirical fits over a wide range of concentrations \cite{EOS}. Then we generalize the equation of state of a monodisperse polymer solution to a simple polydisperse case, assumed to be composed of only two sizes of the same type of polymers: small and big PEGs, or sPEG and bPEG, respectively. The third explicit component of the solution is  water. We will first analyze the osmotic pressure of this solution for various amounts of sPEG added to the background of bPEG. Then we assume that this solution is in equilibrium with a pore that can be penetrated by the sPEG at a finite free energy price but is impenetrable to bPEG. We calculate the corresponding partition coefficient and assess the pushing forces exerted by the external polymer solution {\sl via} its osmotic pressure.

\section{Free energy of a binary polymer mixture}

We start with the free energy $\Delta F(\phi_p)$ of a monodisperse PEG polymer solution as a function of the monomer fraction $\phi_p$ of polymers derived in \cite{Muthu1}, whose range of validity is restricted to long polymer chains such that $\phi_p > {\phi_p}^*$; ${\phi_p}^*$ corresponds to the overlap concentration defining the onset of the semi dilute region, with monomer fraction low enough that higher order terms, above the second virial term, need not be taken into account. This form of the free energy and the consequent osmotic pressure is completely consistent with the phenomenological form of the PEG equation of state demonstrated in \cite{EOS}. 

Under the stated constraintshe the free energy of this monodisperse polymer solution assumes the form \cite{Muthu1}
\begin{eqnarray}
\frac{\Delta F(\phi_p)}{k_B T} &=& n_0 \left( \frac{\phi_p}{N_p}  \ln{\phi_p} + (1-\phi_p) \ln{(1-\phi_p)}  + \chi \phi_p - 
{\textstyle\frac12} \phi_p^{2} + ~\alpha  \left( {\textstyle\frac12} - \chi\right)^{3/4} {\phi_p}^{9/4}\right),
\label{ansatz1}
\end{eqnarray}
where $k_B T$ is the thermal energy. The parameter $\alpha$ in the above free energy expression was evaluated in \cite{Muthu1} as $\alpha = 1.87$; $\chi$ is the Flory-Huggins  parameter and $\phi_p$ is the monomer fraction of the polymer $N_p$ segments long $$\phi_p = \frac{n_p N_p}{n_w + n_pN_p} = \frac{n_p N_p}{n_0},$$where $n_p$ is the number of polymer monomers, $n_w$ is the number of water molecules so that $n_0$ is the total number of monomers plus water molecules. 

We now use this form of the free energy for a different problem, a binary polymer mixture. We assume that each of the two components separately conforms to the limits of validity of the above formula. If the mixture is composed of  $n_s$ molecules of the small polymer $N_s$ monomers long, and $n_b$ molecules of the big polymer $N_b$ monomers long, in an aqueous solvent of $n_w$ water molecules, than the free energy of this polymer mixture is given by 
\begin{eqnarray}
\frac{\Delta F(n_w, n_s, n_b)}{k_B T} &=& n_w  \ln{\phi_w} + n_s  \ln{\phi_s} + n_b  \ln{\phi_b} + \chi n_0 \left( \phi_s + \phi_b\right) - 
{\textstyle\frac12} n_0 \left( \phi_s + \phi_b \right)^{2} + \nonumber\\
& & + ~\alpha n_0 \left( {\textstyle\frac12} - \chi\right)^{3/4} \left( \phi_s + \phi_b \right)^{9/4},
\label{ansatz}
\end{eqnarray}
a direct generalization of Eq. \ref{ansatz1}. The definition of the monomer fractions is
\begin{equation}
\phi_w = \frac{n_w}{n_0} \qquad \phi_s = \frac{n_s N_s}{n_0} \qquad \phi_b = \frac{n_b N_b}{n_0} \qquad {\rm with} \qquad n_0 = n_w + n_s N_s + n_b N_b ,
\end{equation}
so that $\phi_w + \phi_s + \phi_b = 1.$ $n_0$ is the total number of molecules, {\sl i.e.}, the number of water molecules and all the monomers;  the monomer fractions are expressed for long (b) or short (s) polymer molecules.

\section{Osmotic pressure}

The equation of state of the binary polymer mixture is its osmotic pressure as a function of the monomer fractions of the polymer types. We calculate this osmotic pressure $\Pi(\phi_s, \phi_b)$ by first evaluating the chemical potential of water, for given $n_s $ and $n_b $, 
\begin{eqnarray}
\mu_w = - \frac{\overline V \Pi(\phi_s, \phi_b)}{k_B T} = { \frac{\partial }{\partial n_w} \left( \frac{\Delta F}{k_B T}\right)}\vert_{n_s, n_b}.  
\label{potwater}
\end{eqnarray}
where $\overline V $ is the molecular volume of the solvent. The osmotic pressure of the solution in dimensionless form, $\tilde\Pi(\phi_s, \phi_b)$,  is then obtained as 
\begin{eqnarray}
\tilde\Pi(\phi_s, \phi_b) &=& \frac{\overline V \Pi(\phi_s, \phi_b)}{k_B T} = \nonumber\\
&=&-  \ln{\phi_w} + \phi_w -1 + \frac{\phi_s}{N_s} + \frac{\phi_b}{N_b}  - {\textstyle\frac12}\left(1-\phi_w \right)^{2} + {\textstyle\frac{5}{4}} \alpha \left( {\textstyle\frac12} - \chi\right)^{3/4}  \left( 1-\phi_w\right)^{9/4}. 
\label{plothis2}
\end{eqnarray}

\begin{figure*}[]
\begin{center}
		\includegraphics[width= 0.45 \textwidth]{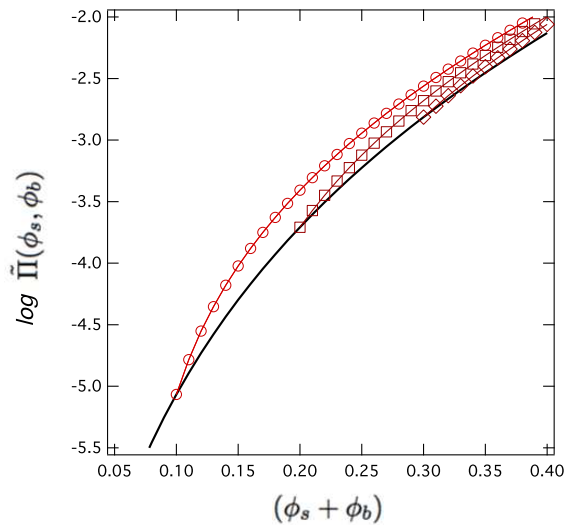} 
\caption{(Color online) Mixture of PEG400 ($N_s \sim 9$) and PEG 3500 ($N_b \sim 40$). Plot of the log of osmotic pressure Eq. \ref{plothis2} as a function of the total polymer monomer fraction (log-linear plot). Four different values of $\phi_s$ are presented, {\sl viz.} $\phi_s =$ 0 (full line), 0.1 ($\Circle$), 0.2 ($\Box$) and 0.3 ($\Diamond$). \label{osmpress}}
\end{center}
\end{figure*}

Remarkably, to the lowest order in $\phi_s$ and $\phi_b$ this dimensionless osmotic pressure is a sum of the (ideal) van't Hoff and the des Cloizeaux terms.
In the limit of a single type of polymer, {\sl i.e.} $\phi_s = 0$ or  $\phi_b = 0$, this decomposition is consistent with the formula that was used to fit the equation of state experimental data in Refs. \cite{EOS1,EOS}. The only difference is in the notation: $\alpha \longrightarrow \alpha \left( {\textstyle\frac12} - \chi\right)^{3/4}$. Introducing $\tilde\alpha = \alpha \left( {\textstyle\frac12} - \chi\right)^{3/4}$ one can deduce that  $\tilde\alpha = 0.49$ for PEG \cite{EOS}.  

Fig. \ref{osmpress} shows the osmotic pressure $\tilde\Pi(\phi_s, \phi_b)$ of a binary mixture of two PEG polymers of different sizes, PEG400  with $N_s \sim 9$ and PEG 3500 with $N_b \sim 40$, obtained from Eq. \ref{plothis2}.
For water ${\overline V} = 1.0 ~\rm ml/g$;  we will assume the same value of molecular volume for PEG  \footnote{In reality for PEG ${\overline V} = 0.825 ml/g$ }. We compare the osmotic pressure for 10, 20 and 30 \% of PEG 400 on top of the background PEG 3500 polymer solution. As seen in Fig. \ref{osmpress}, the osmotic pressure is higher for mixtures with the relative change decreasing with $\phi_b$. It must be noted that the effect of disparity in the molecular weights of the small and the big polymer, {\sl i.e.}, in $N_s$ and $N_b$, appears as the ratio of the total number of big molecules vs. the number of the small molecules, as is evident from Eq. \ref{plothis2}.

\section{Chemical potentials of polymers}

For a mixed PEG solution in equilibrium with a pore that is permeable only to one component of the PEG mixture ({\sl e.g.} sPEG) but impermeable to the other one (bPEG), the short PEG chain has to be in chemical equilibrium between the solution and the pore \cite{Teraoka}. This signifies that the chemical potential of sPEG chain inside the pore and in the external solution must be the same. Thus it makes sense to define the {\sl partition coefficient} of the PEG type that can penetrate the pore as a ratio between the concentration inside the pore and outside in the external solution.

We first derive the chemical potential for the two PEG components of the solution as well as for water. For polymer $s$ or $b$  
\begin{eqnarray}
\mu_{s,b} &=& \frac{\partial }{\partial n_{s,b} } \left( \frac{\Delta F}{k_B T}\right)  =  \ln{\phi_{s,b}} + 1 - \phi_{s,b} - \phi_w N_{s,b} - \phi_{b,s} \frac{N_{s,b}}{N_{b,s}} - \nonumber\\
& &+ (\chi - {\textstyle\frac12}) N_{s,b} + {\textstyle\frac12}N_{s,b }\phi_w^2  -  {\textstyle\frac{5}{4}} \tilde\alpha N_{s,b} \left( \phi_s + \phi_b\right)^{9/4}  +  \frac{9}{4} \tilde\alpha N_{s,b} \left( \phi_s + \phi_b\right)^{5/4}  ,
\label{mu1}
\end{eqnarray}
where, again, all the other variables are kept fixed in taking the derivatives. Using the expression for the chemical potential of water, Eq. \ref{potwater}, we derive
\begin{eqnarray}
\mu_{s,b} - N_{s,b} \mu_w &=&  \ln{\phi_{s,b}} + 1 - N_{s,b} (\ln{\phi_w} +1 - \phi_w) + (\chi - 1) N_{s,b} + N_{s,b} \frac{9}{4} \tilde\alpha \left( \phi_s + \phi_b\right)^{5/4}.
\label{ghjferw2}
\end{eqnarray}
On the l.h.s. of the above equation we have an excess chemical potential of the polymer compared with the same number of solvent molecules as there are monomers. It is easy to show that the chemical potentials derived above satisfy the Gibbs-Duhem relation.

\section{Partition coefficient}

We will now calculate the partition coefficient for small polymers in solution assuming that they can penetrate the pore, schematically depicted in Fig. \ref{schematic}. 

Assume first a polymer solution composed of only small (s) polymers and  that sPEG can enter the pore with a free energy penalty $\Delta f$. For the time being we surmise that this free energy difference does not depend on the state of the polymer in the pore and is independent of the polymer concentration inside the pore, depending only on the single chain confinement free energy. For the magnitude of $\Delta f$, we use the standard argument \cite{Teraoka,Muthubook} based on the fact that the polymer needs to be squeezed into a pore whose diameter is smaller than the "natural" Flory radius of the chain. With this in mind we obtain\cite{Teraoka,Muthubook}
\begin{equation}
\Delta f = N_s \left( \frac{a}{R}\right)^{\frac53} = f_0(R) N_s,
\end{equation}
where $a$ is the monomer size and where we assume that the length of the polymer that penetrates the pore is equal to the full length of polymer, $N_s$. In what follows we will simply take the value of  $\Delta f$ to be independent of $\phi_{s,b}$ and do not specify its exact form. The latter indeed depends not only on the pore radius but also on the polymer  - pore surface interaction energy, for details see Ref. \cite{Muthubook}. 

\begin{figure}[t!]
\centering
\includegraphics[width= 0.45 \textwidth]{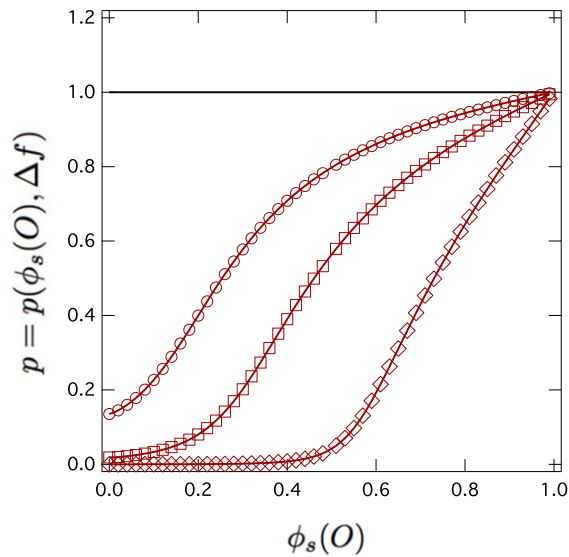}
\caption{Plot of the partition coefficient Eq. \ref{defp2} for a single species of polymer, PEG400 with $N_s = 9$, as a function of the polymer monomer fraction $\phi_s(O)$ in the external solution. Cases with four different penetration free energies are presented, $\Delta f =$ 0 (full line), 2 ($\Circle$), 4 ($\Box$), 9 ($\Diamond$) in dimensionless units.\label{part-single}}
\end{figure}

Then if $I$ stands for the inside the pore and $O$ for the outside, the chemical equilibrium is established when 
\begin{equation}
\mu_s (I) + \Delta f = \mu_s(O).
\label{grapel1}
\end{equation}
Also, since the solvent is in chemical equilibrium too, we need an additional equation
\begin{equation}
\mu_w (I) = \mu_w(O).
\end{equation}
In principle at least, for the solvent too we could add a separate pore penetration energy. For now we do not explore this scenario. The above two equations can then be rewritten in an equivalent form
\begin{equation}
\mu_{s}(I) - N_s \mu_w(I) + \Delta f = \mu_{s}(O) - N_s \mu_w(O),
\end{equation}
where we took into account the assumption that all the monomers and water molecules are of the same size.
Taking into account Eq. \ref{mu1}
\begin{eqnarray}
& & \ln{\phi_{s}(I)} - N_s (\ln{\phi_w(I)} +1 - \phi_w(I)) + N_s \frac{9}{4} \tilde\alpha \left( \phi_s(I) + \phi_b(I)\right)^{5/4} + \Delta f = \nonumber\\
&& \ln{\phi_{s}(O)} - N_s (\ln{\phi_w(O)} +1 - \phi_w(O)) + N_s \frac{9}{4} \tilde\alpha \left( \phi_s(O) + \phi_b(O)\right)^{5/4},
\label{ghewq}
\end{eqnarray}
where we have removed all the irrelevant constants. Equation \ref{ghewq} takes into account the chemical equilibrium of the pore-external solution system in terms of the polymer as well the solvent chemical potentials. Since, by assumption, the big polymer can not penetrate the pore we obviously have $ \phi_b(I) = 0$.  Introducing now the partition coefficient 
$$p = {\frac{\phi_s(I)}{\phi_s(O)}},$$Eq. \ref{ghewq} assumes the final form 
\begin{eqnarray}
\ln{p} + \Delta f &=& N_s \left( \ln\frac{(1 - p\phi_s(O))}{(1 - \phi_s(O) - \phi_b(O))} + (p\phi_{s}(O) - \phi_s(O) - \phi_b(O))+ \right.\nonumber\\
&& \left. + \frac{9}{4} \tilde\alpha \left( \left( \phi_s(O) + \phi_b(O)\right)^{5/4} - (p\phi_s(O))^{5/4} \right)\right).
\label{defp1}
\end{eqnarray}
The solution of this equation gives us $p = p(\phi_s(O), \Delta f)$ when  solved numerically. Usually $ \Delta f$ is assumed to be a linear function of $N_s$ but this is irrelevant here. 

\section{Partition coefficient: single polymer type}

Consider first that we have only small polymer, e.g. PEG400, in the pore as well as in the external solution, {\sl i.e.} $\phi_b(O) = 0$. In this case, Eq. \ref{defp1} reduces to 
\begin{equation}
\ln{p} + \Delta f = N_s \left( \ln\frac{(1 - p\phi_s(O))}{(1 - \phi_s(O))} + (p - 1) \phi_{s}(O) + \frac{9}{4} \tilde\alpha \left(1 - p^{5/4} \right)\phi_s(O)^{5/4}\right).
\label{defp2}
\end{equation}
Fig. \ref{part-single} shows numerical solution assuming that $\tilde\alpha = 0.49$ for PEG400. We evaluate the partition coefficient $$p = p(\phi_s(O), \Delta f)$$and plot it for different $\Delta f$ as a function of $\phi_s(O)$ in the external solution.  

The general features of the polymer partitioning in this case are well known: the partition coefficient is a monotonically increasing function of the monomer fraction of the polymer in the external solution, $\phi_s(O)$, and strictly $p \leq 1$, {\sl i.e.} the monomer fraction inside the pore is always smaller than outside. The additional free energy penalty, $\Delta f $, from polymer-pore interaction, makes it clearly more difficult for the polymer to enter the pore. Consequently higher values of the monomer fraction in the external solution are needed to reach the same partition coefficient. Not only that, as the penetration free energy penalty grows, the partition coefficient becomes a steeper function of the monomer fraction in the external solution, with a narrower interval of saturation behavior.

\begin{figure*}[]
\begin{center}
\begin{minipage}[b]{8.5 cm}\begin{center}
\includegraphics[width=\textwidth]{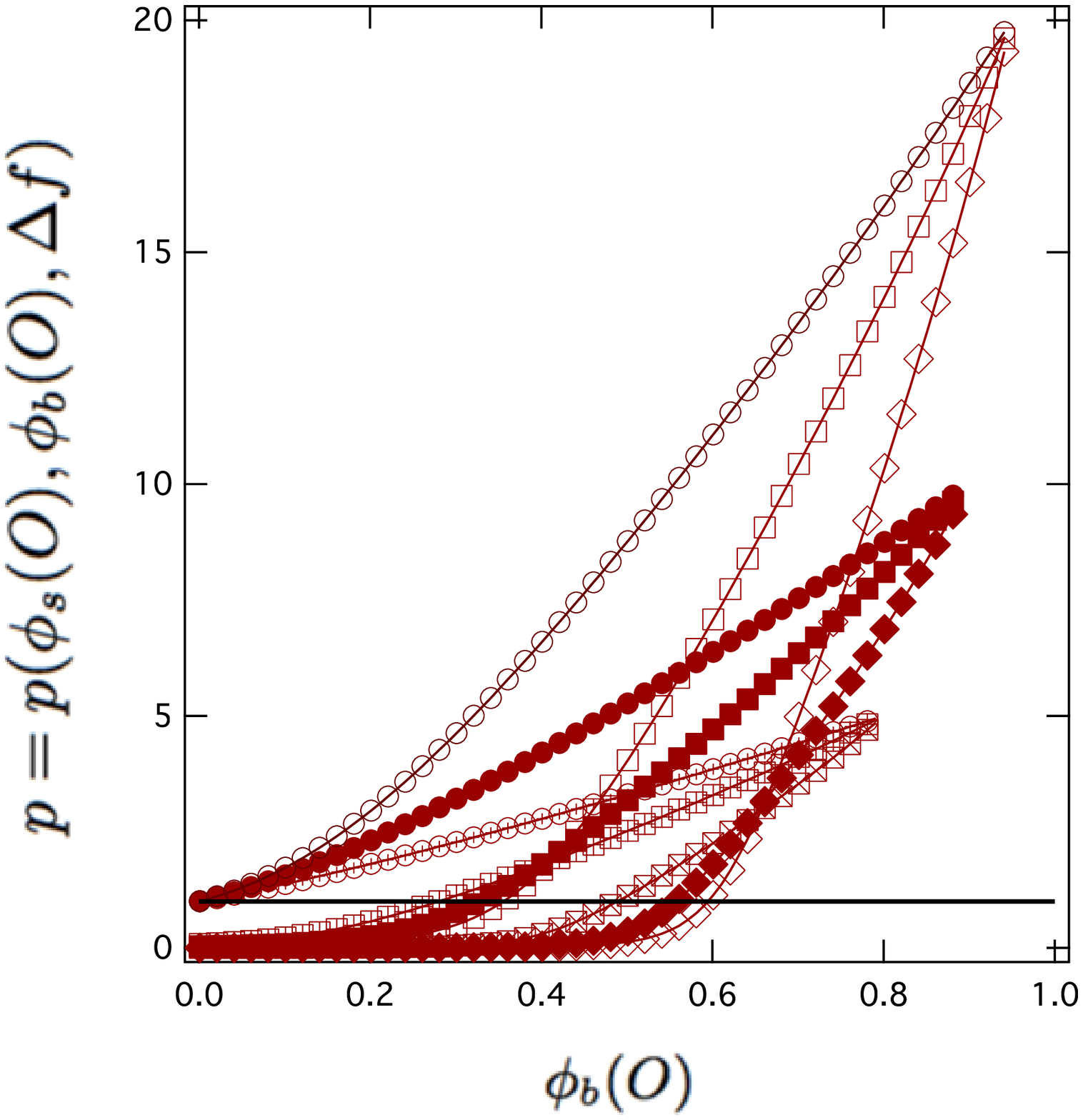} (a)
	\end{center}\end{minipage}  \hskip0.25cm	
	\begin{minipage}[b]{8.6 cm}\begin{center}
\includegraphics[width=\textwidth]{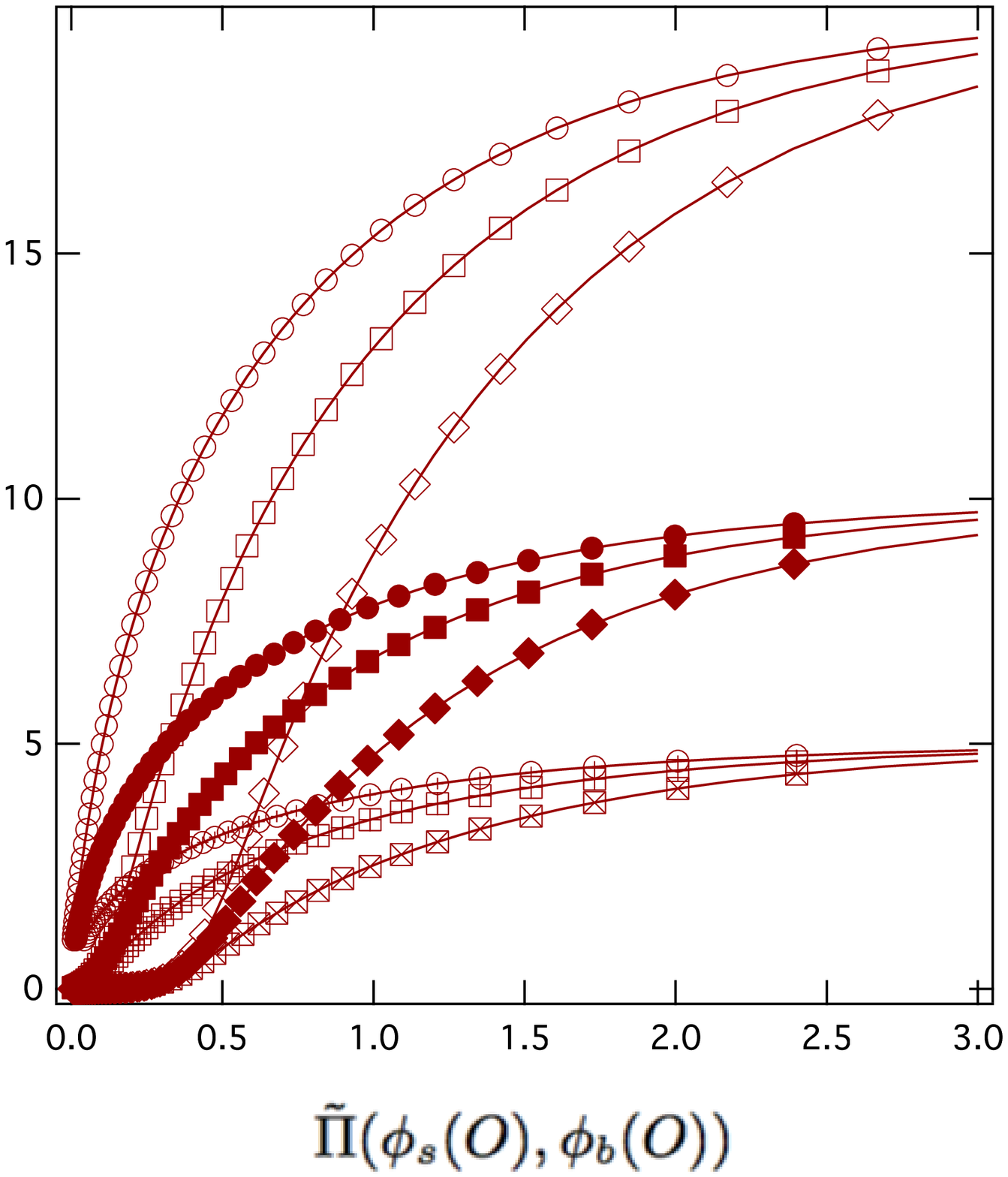} (b)
	\end{center}\end{minipage}
	\begin{minipage}[b]{8.5 cm}\begin{center}
\includegraphics[width=\textwidth]{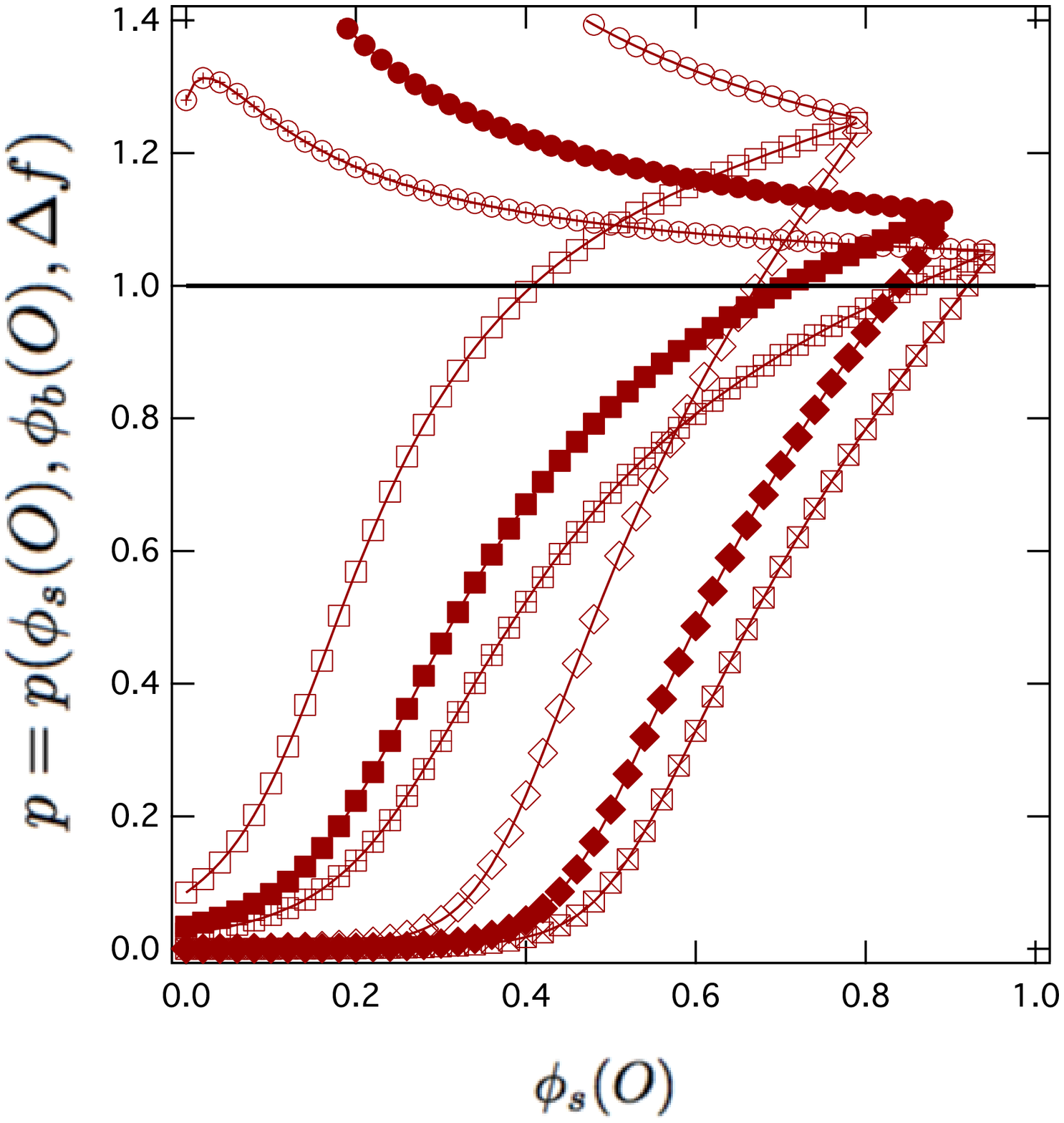} (c)
	\end{center}\end{minipage}  \hskip0.25cm
	\begin{minipage}[b]{8.6 cm}\begin{center}
		\includegraphics[width=\textwidth]{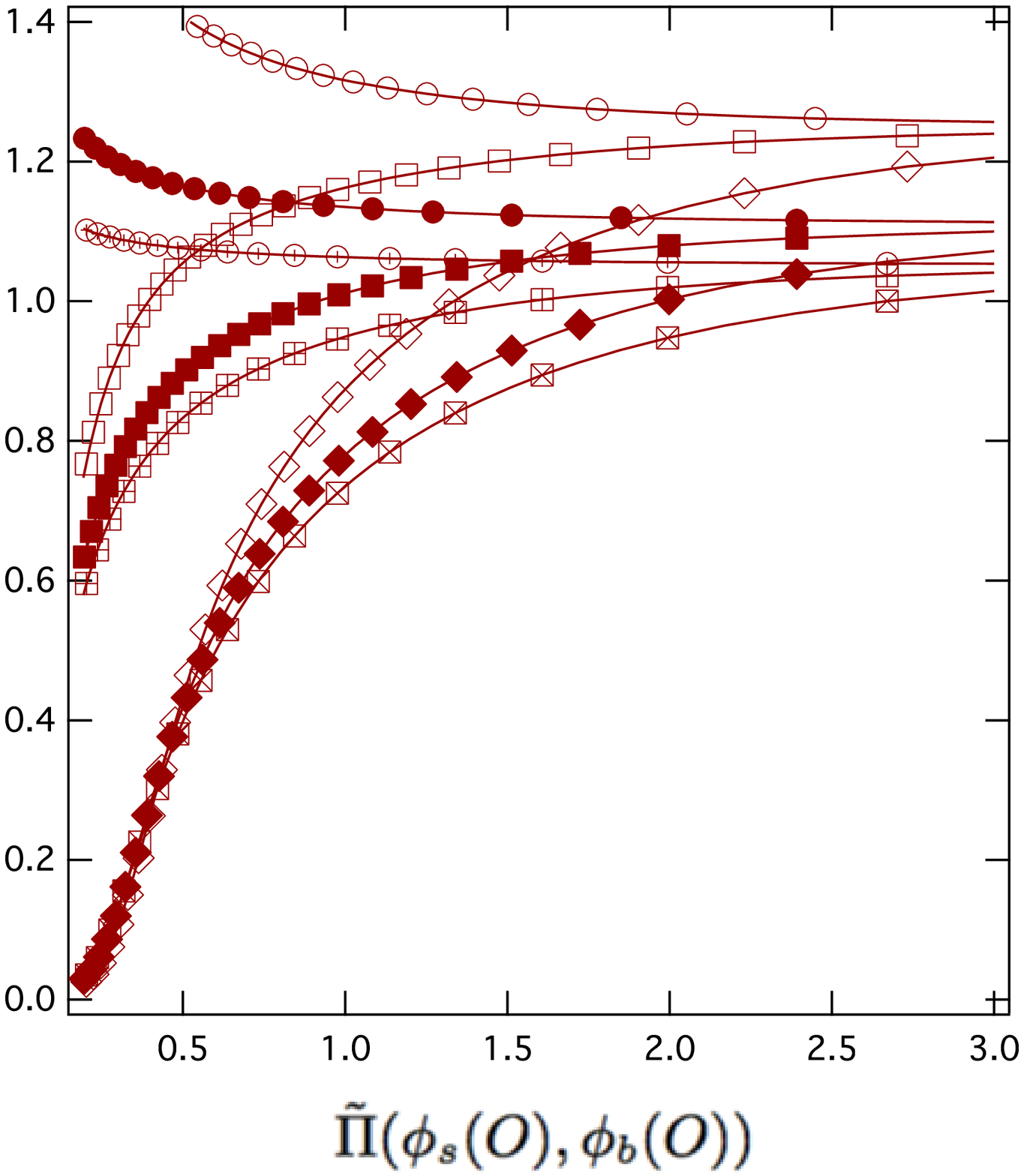} (d)
	\end{center}\end{minipage} 
	\vskip 0.25cm	 
\caption{(Color online) (a) Partition coefficient Eq. \ref{defp3} for different values of $\phi_b(O)$ and different values of the additional polymer-pore interaction free energy $\Delta f$, at a fixed value of $\phi_s(O)$ in the external solution. $\phi_s(O) = 0.05$ and  $\Delta f = 0, 4, 9$ ($\Circle$, $\Box$, $\Diamond$). $\phi_s(O) = 0.1$ and  $\Delta f = 0, 4, 9$ ($\CIRCLE$, $\blacksquare$, $\blacklozenge$). $\phi_s(O) = 0.2$ and  $\Delta f = 0, 4, 9$ ($\oplus$, $\boxplus$, $\boxtimes$). The delimiting value $p = 1$ is shown as a black line.   \\ (b) Partition coefficient for different values of $\phi_b(O)$ and the same fixed value of $\phi_s(O)$ in external solution, as a function of total dimensionless osmotic pressure $\tilde\Pi(\phi_s(O), \phi_b(O))$. Numerical results for three different values of $\Delta f$ and three different values of $\phi_s(O)$  are presented. \\ (c) Partition coefficient $p( \phi_s(O), \phi_b(O), \Delta f)$,  Eq. \ref{defp3}, for different values of PEG400 ($N_b = 9$) monomer fractions $\phi_s(O)$ and different values of $\Delta f$, at a fixed value of PEG3500 ($N_b = 40$) $\phi_b$ in the external solution. $\phi_b(O) = 0.2$ and  $\Delta f = 0, 4, 9$ ($\Circle$, $\Box$, $\Diamond$). $\phi_b(O) = 0.1$ and  $\Delta f = 0, 4, 9$ ($\CIRCLE$, $\blacksquare$, $\blacklozenge$). $\phi_b(O) = 0.05$ and  $\Delta f = 0, 4, 9$ ($\oplus$, $\boxplus$, $\boxtimes$). The delimiting value $p = 1$ is again shown as a black line.  \\ (d) Partition coefficient for different values of $\phi_s(O)$ at the same fixed values of $\phi_b(O)$ in external solution, as a function of the total dimensionless osmotic pressure $\tilde\Pi(\phi_s(O), \phi_b(O))$. Numerical results for three different values of $\Delta f$ are presented and three different values of $\phi_b(O)$.  \label{fig:SCimp1}}
\end{center}
\end{figure*}

We now evaluate explicitly two limiting cases for $\phi_{s}(O)  \ll 1$ and $\phi_{s}(O)  \rightarrow 1$. In the first case $\phi_{s}(I) = p \phi_{s}(O) < \phi_{s}(O)$ so that obviously $p \ll 1$, we obtain to the lowest order
\begin{equation}
\ln{p} + \Delta f = N_s \left( \ul{9}{4} \tilde\alpha\left(1-p^{5/4}\right) \phi_{s}(O) ^{5/4}+ {\cal O}[\phi_{s}(O)  ]^{2}\right) \simeq N_s  \ul{9}{4} \tilde\alpha ~\phi_{s}(O) ^{5/4}.
\label{defp3}
\end{equation}
and therefore
\begin{equation}
\ln{p(\phi_{s}(O) )}  \simeq -\Delta f  + N_s \ul{9}{4} \tilde\alpha ~\phi_{s}(O) ^{5/4}.
\label{defp3}
\end{equation}
The partition coefficient is thus increased from its ideal value given by $\ln{p(\phi_{s}(O) )} = \ln{p_0} = -\Delta f $. In the opposite limit of $\phi_{s}(O), p  \rightarrow 1$, within the range of $\vert p - 1\vert \ll \Delta f$, 
\begin{equation}
\ln{p(\phi_{s}(O) )} \simeq -  {\Delta f}{\left( N_s \ul{9}{4} \tilde\alpha ~\phi_{s}(O) ^{5/4}\right)^{-1}},
\label{defp5}
\end{equation}
{\sl i.e.}, the partition coefficient is decreased from its ideal value. These two limiting cases can be well discerned also from the complete numerical solution of Eq. \ref{defp2} (Fig. \ref{part-single}). It is clear from this figure that the range of validity of the limiting expression Eq. \ref{defp5} depends on the value of $\Delta f$.

\section{Partition coefficient: two polymer types}

We now return to the original problem of a polymer mixture, in the simplest case composed of two types of polymer, $b$ and $s$, of which only the $s$ can penetrate the pore. In this case, we need to solve the full equation Eq. \ref{defp1}. The solution
\begin{equation}
p = p( \phi_s(O), \phi_b(O), \Delta f),
\end{equation}
is a function of the monomer fraction of the small and the big polymer in the external solution as well as of the free energy penalty for entering the pore. We plot the solution $p = p( \phi_s(O), \phi_b(O))$ for different values of $\Delta f$ as a parameter, which corresponds to the variation of the penetration free energy with the pore size. As an elucidating example, we again have for "s" polymer PEG400 with $\rm N_s = 9$ and for "b" polymer, PEG 3500 with $N_b = 40$. We study separately the dependence of the partition coefficient on $\phi_b(O)$ with $\phi_s(O)$ fixed, and {\sl vice versa}. Both cases differ fundamentally from the single polymer type pore partitioning.

\begin{figure}[t!]
	\begin{minipage}[b]{7.9 cm}\begin{center}
\includegraphics[width=\textwidth]{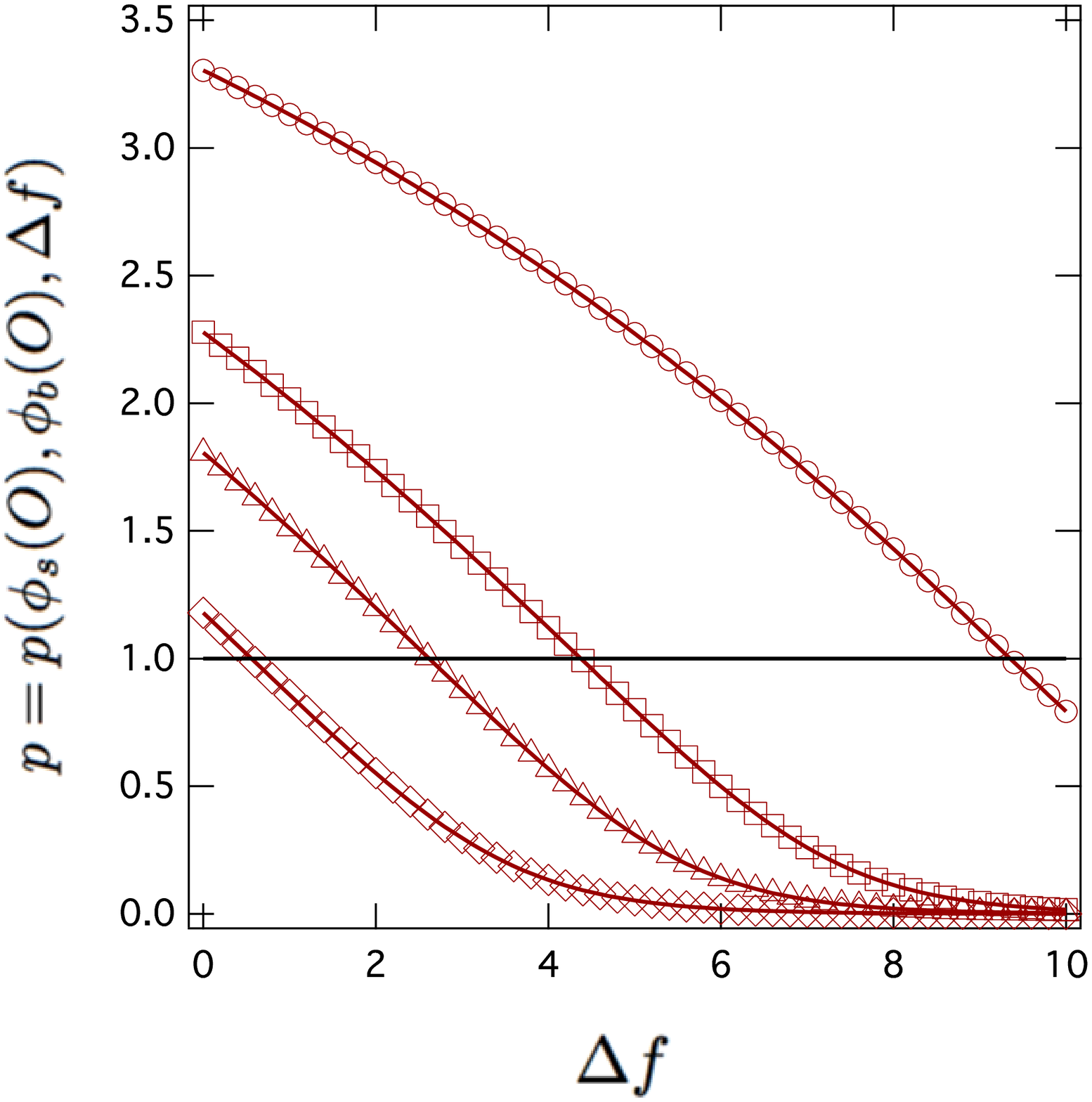} (a)
	\end{center}\end{minipage}  \hskip0.25cm
	\begin{minipage}[b]{8.6 cm}\begin{center}
		\includegraphics[width=\textwidth]{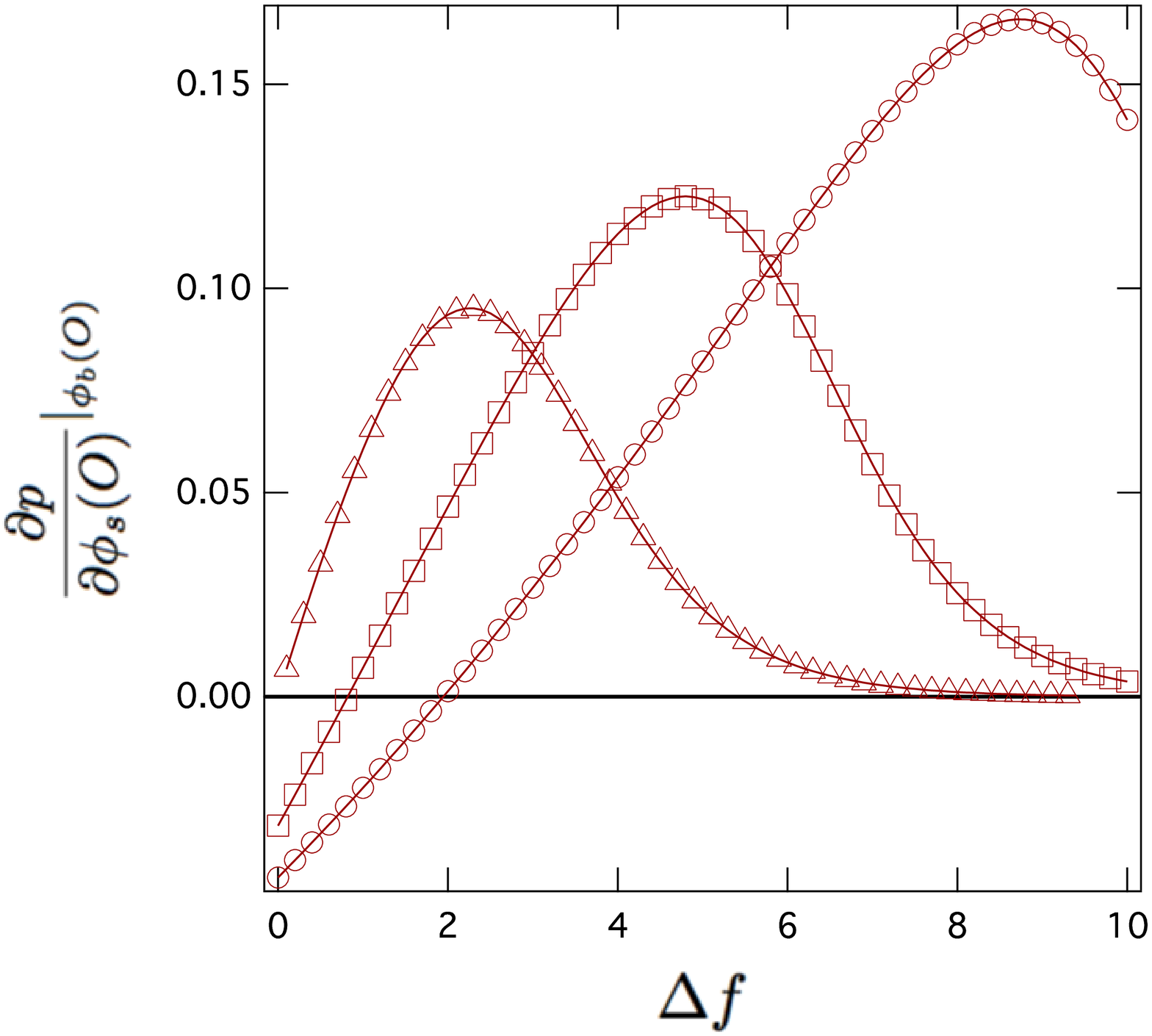} (b)
	\end{center}\end{minipage} 
	\vskip 0.25cm
\caption{(a) Partition coefficient Eq. \ref{defp2} for PEG400 ($N_s = 9$) as a function of $\Delta f$, for $\phi_s(O)=0.2$, $\phi_b(O) = 0.5$ ($\Circle$), $\phi_s(O)=0.2$, $\phi_b(O) = 0.3$ ($\Box$), $\phi_s(O)=0.2$, $\phi_b(O) = 0.2$ ($\triangle$) and $\phi_s(O)=0.2$, $\phi_b(O) = 0.05$ ($\Diamond$). Partition coefficient thus appears as a monotonically decreasing function of the pore penetration energy $\Delta f$. (b) Derivative of the partition coefficient $ \frac{\partial{p}}{\partial{\phi_s(O)}}{|_{\phi_b(O)}}$ from Eq. \ref{defder} as a function of $\Delta f$, for different external solution compositions: $\phi_s(O) = 0.2$, $\phi_b(O) = 0$ ($\triangle$), $\phi_s(O) = 0.3$, $\phi_b(O) = 0.1$ ($\Box$), $\phi_s(O) = 0.4$, $\phi_b(O) = 0.2$ ($\Circle$). The derivative changes sign at a "critical point" of the pore penetration free energy,  $\Delta f_c \simeq 1$ in the second case and $\Delta f_c \simeq 2$ in the third case. For large enough $\Delta f$ the derivative is positive and limits to zero. \label{dependence-on-f}}
\end{figure}

Fig. \ref{fig:SCimp1} shows the dependence of the partition coefficient on the composition of the external solution, $\phi_s(O), \phi_b(O)$, as well as the free energy penalty $\Delta f$. Consider first the dependence of the partition coefficient on $\phi_b(O)$ at fixed $\phi_s(O)$, Fig. \ref{fig:SCimp1}(a), (b). We see that at every set $\phi_s(O)$ the addition of the non-penetrating polymer "b" to the external solution monotonically increases the partition coefficient of the penetrating one, (Fig. \ref{fig:SCimp1}(a)). We can therefore conclude that the "b" polymer always pushes the "s" polymer into the pore. This is the physical meaning of the "polymers pushing polymers" concept.  In addition, for every composition, {\sl i.e.}, $\phi_s(O), \phi_b(O)$,  there exists a "critical value" of the penetration free energy penalty at which the monotonically decreasing partition coefficient crosses the value $p = 1$, see Fig. \ref{dependence-on-f}(a), defined as $\Delta f_1(\phi_s(O), \phi_b(O)) = \Delta f (p = 1)$, where we indicated explicitly that it depends on the composition of the external polymer solution.  Depending on the composition of the external solution and for $\Delta f < \Delta f_1$, we can have enhanced partitioning of the penetrating polymer with $p > 1$. This is very different from the case of a single polymer type where always $p \leq1$. Note also that the derivative of the partition coefficient with respect to $\phi_b(O)$ is always positive, for any composition and any penetration penalty. The partition coefficient can be analyzed alternatively also through its dependence on the total osmotic pressure of the external solution, obtained from Eq. \ref{plothis2}, (Fig. \ref{fig:SCimp1}(b)). Clearly for fixed $\phi_s(O)$ the polymer "s" is always pushed into the pore, Fig. \ref{fig:SCimp1}(b), as its partition coefficient is an increasing function of the total external solution osmotic pressure. 

\begin{figure}[t!]
	\begin{minipage}[b]{8.6 cm}\begin{center}
\includegraphics[width=\textwidth]{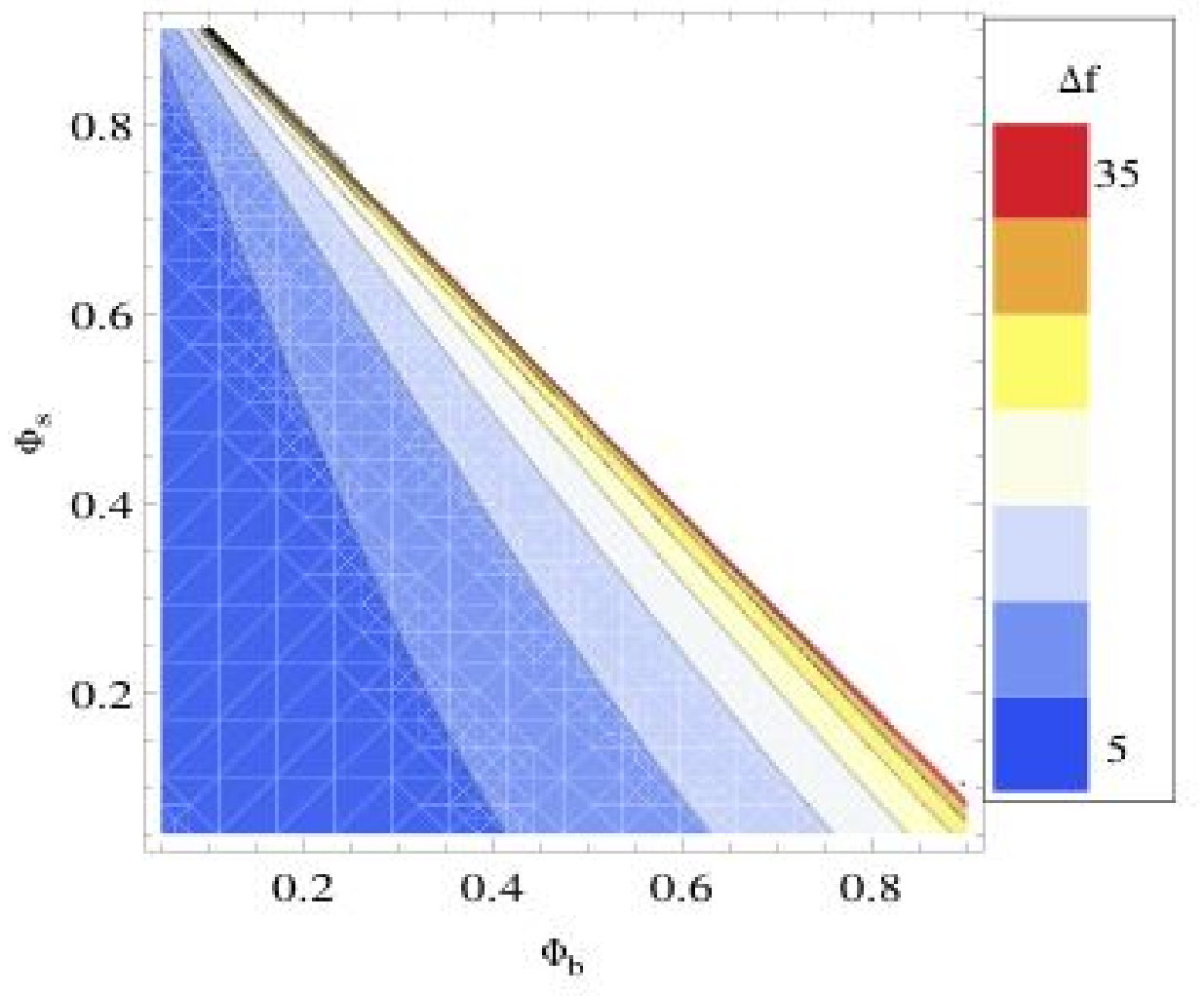} (a)
	\end{center}\end{minipage}  \hskip0.25cm
	\begin{minipage}[b]{8.6 cm}\begin{center}
		\includegraphics[width=\textwidth]{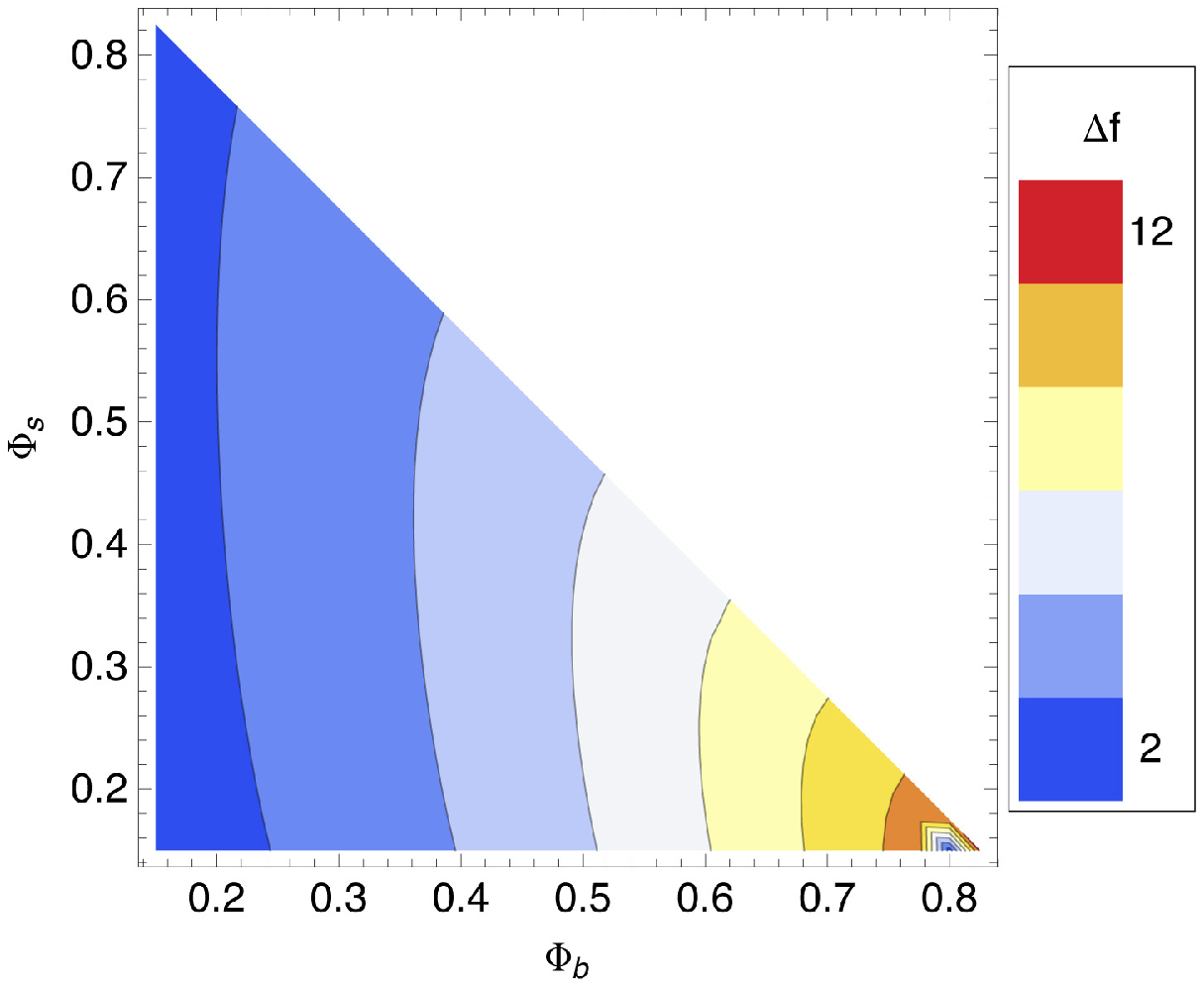} (b)
	\end{center}\end{minipage} 
	\vskip 0.25cm
\caption{(a) "Critical value" of the penetration free energy penalty defined as $\Delta f_1(\phi_s(O), \phi_b(O)) = \Delta f (p = 1)$ for each particular composition of the binary mixture $\phi_s(O), \phi_b(O)$. As long as $\Delta f < \Delta f_1$ there is enhanced partitioning or {\sl superpartitioning} of the penetrating polymer with $p > 1$. $\Delta f_1(\phi_s(O), \phi_b(O)) $ is a monotonic function of the composition. (b) "Critical value" of the penetration free energy penalty defined as $\Delta f_0(\phi_s(O), \phi_b(O)) = \Delta f (p' = 0)$. For $\Delta f < \Delta f_0$ we have $p' = \frac{\partial{p}}{\partial{\phi_s(O)}} < 0$ and the opposite for $\Delta f > \Delta f_0$, thus leading to augmented and depleted partitioning regions in the parameter space. $\Delta f_0(\phi_s(O), \phi_b(O)) $ is obviously a non-monotonic function of the composition. \label{critical-f}}
\end{figure}

The converse case of the dependence of the partition coefficient on $\phi_s(O)$ with constant $\phi_b(O)$ is more complicated. First of all, in this case too we discern, (Fig. \ref{fig:SCimp1}(c)), that polymer "s" can have enhanced partitioning, {\sl i.e.}, its partition coefficient can be larger than one. The corresponding "critical value" of the penetration free energy penalty is the same as in the previous case, $\Delta f_1(\phi_s(O), \phi_b(O)) = \Delta f (p = 1)$ and is an increasing function of both $\phi_s(O)$ and $\phi_b(O)$, but is not symmetric in the two arguments, as can be clearly seen from Fig. \ref{critical-f}(a). There is, however, an additional feature in the $\phi_s(O)$ variation of the partition coefficient when $\phi_b(O)$ is kept fixed. From Fig. \ref{fig:SCimp1}(c), we discern that for any non-zero value of $\phi_b(O)$, there is a range of $\Delta f$ values where the rate of variation of the partition coefficient with respect to $\phi_s(O)$, {\sl i.e.}, $p' \equiv \frac{\partial{p}}{\partial{\phi_s(O)}}$, changes sign. Apart from possible enhanced partitioning we can thus also have an augmented, $p' > 0$, and a depleted rate of partitioning with $p' < 0$. In complete analogy with the behavior of the partition coefficient itself, we can define a "critical value" of the penetration free energy as $\Delta f_0(\phi_s(O), \phi_b(O)) = \Delta f (p' = 0)$, delimiting the augmented and the depleted partitioning regions, see Fig. \ref{dependence-on-f}(b) and Fig. \ref{critical-f}(b).  

Fig. \ref{critical-f}(b) also shows that  $\Delta f_0(\phi_s(O), \phi_b(O))$ is not a monotonic function of its arguments. If we fix $ \phi_b(O)$ and follow the variation of $\Delta f_0$ with $\phi_s(O)$, we note that it first increases then decreases, while it appears to be strictly monotonic in the converse case of fixed $\phi_s(O)$ and variable $\phi_b(O)$. The non-monotonic variation of $\Delta f_0(\phi_s(O), \phi_b(O))$ indicates a complicated interplay of various parts of the free energy on polymer penetration of the pore, specifically of the polymer mixing entropy and configurational fluctuations in the external binary polymer solution and the penetration penalty for entering the pore by only one type of the polymers. There is no analogous effect in the case of a single polymer type inside and outside. 

Since the depleted partitioning of polymer "s" form the pore, as defined above, is a counterintuitive phenomenon, it needs some further elucidation. For every composition of the system, that is for every fixed pair $\phi_s(O), \phi_b(O)$, there is thus a corresponding $\Delta f_0(\phi_s(O), \phi_b(O))$, where the rate of variation of the partition coefficient with respect to $\phi_s(O)$, changes sign. In fact one can derive the following simple relationship 
\begin{equation}
\frac{\partial \log{p}}{\partial \log{\phi_s(O)}}{|_{\phi_b(O)}} = \frac{\phi_s(O)}{p} \frac{\partial{p}}{\partial{\phi_s(O)}}{|_{\phi_b(O)}} = \frac{N_s \left( \chi_s(O) - \chi_s(I)\right)}{1 + N_s \chi_s(I)},
\label{defder} 
\end{equation}
where we introduced the osmotic compressibilities of the penetrating polymer as
\begin{equation}
\chi_s(*) = \phi_s(*) \frac{\partial \tilde\Pi }{\partial \phi_s(*)}.
\end{equation}
Here $\phi_s(* = O, I)$ stands for the monomer fraction outside or inside of the pore, while $ \tilde\Pi =  \tilde\Pi(\phi_s(O,I), \phi_b(O,I)) $ is the corresponding dimensionless osmotic pressure, with both osmotic compressibilities depending implicitly on $\Delta f$. From Eq. \ref{defder} it follows that the rate of variation of the partition coefficient with respect to $\phi_s(O)$ depends on the difference of the osmotic compressibilities of the penetrating polymer outside and inside the pore. The transition from augmented to depleted partitioning is thus connected with the osmotic compressibility of the exchangeable polymer in the two environments and one could say that $\Delta f_0(\phi_s(O), \phi_b(O))$ is defined also by the equality of the two compressibilituies, $\chi_s(O)$ and $\chi_s(I)$.

The behavior of the polymer partitioning in the case of variable $\phi_s(O)$ is therefore governed by two energy scales $\Delta f_0$ and $\Delta f_1$ determining the rate of variation and the magnitude of the partition coefficient $p(\phi_s(O), \phi_b(O), \Delta f)$. It is quite unexpected that the mere introduction of an additional non-penetrating polymer to the external solution would have such fundamental and non-trivial repercussions for the partitioning of the other polymer. 

\section{Discussion and Conclusions}

We present a thermodynamic equilibrium analysis of pore penetration by polymer chains in a binary polymer mixture, where only one of the components is allowed to penetrate the pore with a set free energy penalty. 

First we introduced an equation of state, {\sl i.e.}, the dependence of osmotic pressure on the monomer fraction of the polymers, Eq. \ref{plothis2}, consistent with empirical fits obtained for the most complete set of experimental data available for PEG in aqueous solvent as well as PAMS in toluene \cite{EOS}. The proposed equation of state captures all the features of the empirical fits. We generalized the equation of state for a single type of polymer to the case of a simple binary polymer mixture, composed of two types of polymer chains, "s" and "b", of which only the former can enter the pore with a finite free energy price. We then used this equation of state in order to formulate the equilibrium distribution of the short polymer between the external solution and the pore, obtaining the partition coefficient of the short chains in terms of the external polymer solution composition. Our analysis reduces to the known cases when there is only a single type of polymer in solution.

The partitioning of a single component of the polymer solution into a pore can be described with the concept of "polymers pushing polymers": the non-penetrating polymer chains "b" contribute an osmotic push for an enhanced partitioning of the "s" polymer chains into the pore. We find that the larger the concentration of the non-penetrating "b" polymer in the external solution, the larger the concentration of penetrating polymer "s" that will be pushed into the pore at any value of the pore penetration energy. However, this does not imply that the partitioning {\sl coefficient}, {\sl i.e.}, the ratio of the "s" concentration inside and outside the pore, is always a monotonic function of the external concentrations of both polymer types.

Analysing the general properties of the partition coefficient when the "s" and the "b" components of the external polymer mixture are allowed to vary independently, we find that the general features of the pore penetration are governed by two different penetration free energy scales, $\Delta f_1$ and $\Delta f_0$. The former differentiates between enhanced partitioning, $p > 1$, and ordinary partitioning $p < 1$ dependent on the osmotic pushing strength of the external polymer mixture and indeed embodies the concept of "polymers pushing polymers". The other penetration free energy scale, $\Delta f_0$, that differentiates between the rates of variation of the partitioning coefficient of the penetrating "s" polymer leading to an augmented, $p' > 0$, and a depleted rate of partitioning, $p' < 0$. While $\Delta f_1(\phi_s(O), \phi_b(O))$ is a monotonic function of its arguments, $\Delta f_0(\phi_s(O), \phi_b(O))$ is not. The latter type of behaviour is understood as stemming from the interplay between the polymer mixing entropy and/or configurational fluctuations in the external binary polymer solution, and the pore penetration penalty.

Both of these free energy scales depend on the composition of the external solution and for a fixed $\Delta f$ one should be able to control the partitioning of a polymer into a pore, or in general into any nano-cavity, by varying the composition of the external polymer mixture. This principle should have applications in many areas of nano-science dealing with partitioning of polymer chains into small enclosures. Experiments are under way to test at least some of the conclusions reached on the basis of the theory presented here \cite{Philip}.

While we have analyzed only the most simple case of a binary polymer mixture, one can in principle generalize this theory to more complex polydispersity. This would entail the introduction of a size distribution in the equation of state as well as an energy size distribution for the pore penetration penalty. One could formulate the corresponding expressions for equilibrium partitioning in a parallel fashion to what we derived for the binary mixture. This would open the possibility to address new problems in the theory of confined polymer solutions. Furthermore it is easy to envision cases where the polymer mixture would be actually composed of polymers of very different nature such as, {\sl e.g.}, PEG and proteins, or PEG and sugars. It is reasonable to expect that a similar analysis would apply to those cases too.

\section{Acknowledgments}

We thank A. Lo\v sdorfer Bo\v zi\v c for his help with graphics. Research was supported by the U.S. Department of Energy, Office of Basic Energy Sciences, Division of Materials Sciences and Engineering under Award DE-FG02-84DR45170. MM acknowledges the support of NIH grant no. $\rm R01HG002776-09$.


\begin{thebibliography}{9}
\bibitem{Teraoka} I. Teraoka, {\sl Polymer Solutions: An Introduction to Physical Properties}, John Wiley \& Sons, Inc. (2002). 
\bibitem{Muthubook} M. Muthukumar, {\sl Polymer Translocation}, CRC Press; 1 edition (2011).
\bibitem{OST} V.A. Parsegian, R. P. Rand, N. L. Fuller and D. C. Rau, {\sl Methods Enzymol.} (1986) {\bf 127} 400.
\bibitem{OST1} V.A. Parsegian, R. P. Rand and D. C. Rau, {\sl PNAS} (2000) {\bf 97} 3987.
\bibitem{Krasilnikov} T.K. Rostovtseva, E.M. Nestorovich, S.M.  Bezrukov,  {\sl Biophys. J.} (2002) {\bf 82} 160. 
\bibitem{Krasilnikov2} S. M. Bezrukov,  I. Vodyanoy,  R.A. Brutyan,  J.J. Kasianowicz, {\sl Macromolecules} (1996) {\bf 29} 8517. 
\bibitem{Krasilnikov3} O.V. Krasilnikov and S. M. Bezrukov, {\sl Macromolecules} (2004) {\bf 37} 2650.
\bibitem{J2K} J.W. F. Robertson, C.G. Rodrigues, V.M. Stanford, K.A. Rubinson, O.V. Krasilnikov, and J.J. Kasianowicz, {\sl PNAS} (2007) {\bf 104} 8207. 
\bibitem{J2K2} O.V. Krasilnikov, C.G. Rodrigues, and S.M. Bezrukov, {\sl PRL } (2006) {\bf 97} 018301. 
\bibitem{J2K3}J.E. Reiner, J.J. Kasianowicz, B.J. Nablo, and J.W. F. Robertson, {\sl PNAS } (2010) {\bf 107} 12080. 
\bibitem{Gelbart} P. Prinsen, Li Tai Fang, A. M. Yoffe, C. M. Knobler, and W. M. Gelbart, {\sl J. Phys. Chem. B} (2009) {\bf 113} 3873.
\bibitem{Evilevitch} E. Nurmemmedova, M. Castelnovo, C. Catalano and A. Evilevitch, {\sl Quart. Rev. Biophys.} (2007) {\bf 40} 327.
\bibitem{Muthu1} M. Muthukumar, {\sl J. Chem.Phys.} (1986) {\bf 85} 4722.
\bibitem{Muthu2} M. Muthukumar and S.J. Edwards,  {\sl J. Chem.Phys.} (1982) {\bf 76} 2720.
\bibitem{EOS1} P.L. Hansen, J.A. Cohen, R. Podgornik and V.A. Parsegian, {\sl Biophys. J.} (2003) {\bf 84} 350.
\bibitem{EOS} J.A. Cohen, R. Podgornik, P.L. Hansen, V.A. Parsegian, {\sl J. Phys. Chem.} (2009) {\bf 113}, 3709.
\bibitem{Sergey}  V.Y. Zitserman,  A.M.  Berezhkovskii,  V.A. Parsegian,  S.M. Bezrukov, {\sl J. Chem. Phys.} (2005) {\bf 123}, 146101.
\bibitem{Daoud} M. Daoud and P.-G. de Gennes, J. de Physique  (1977) {\bf 38}, 85.
\bibitem{Philip} P. Gournev, S.M. Bezrukov, V.A. Parsegian (in preparation).
\end{thebibliography}
\end{document}